\documentclass{article}
\usepackage{spconf,amsmath}
\usepackage{amsmath}

\usepackage{math}
\usepackage{hyperref}
\usepackage{cite}
\usepackage{amsmath}
\usepackage{amssymb}
\usepackage{steinmetz}
\usepackage{booktabs} 
\usepackage{graphicx}  
\usepackage{float}
\usepackage{multicol}

\usepackage{footnote}
\makesavenoteenv{tabular}
\makesavenoteenv{table}

\DeclareMathOperator{\sinc}{sinc}
\newcommand{\minus}{\scalebox{0.75}[1.0]{$-$}}
\newcommand{\Test}[1]{\expandafter\hat#1}

\title{Filterbank design for end-to-end speech separation}
\name{Manuel Pariente $^1$, Samuele Cornell $^2$, Antoine Deleforge $^1$, Emmanuel Vincent $^1$\thanks{Experiments presented in this paper were partially carried out using the Grid’5000 testbed, supported by a scientific interest group hosted by Inria and including CNRS, RENATER and several Universities as well as other organizations (see \texttt{https://www.grid5000.fr}).}\thanks{High Performance Computing resources were partially provided by the EXPLOR centre hosted by the University de Lorraine}\thanks{The work reported here was started at JSALT 2019, and supported by JHU with gifts from Amazon, Facebook, Google, and Microsoft.}}
\address{$^1$Universit\'e de Lorraine, CNRS, Inria, LORIA, F-54000 Nancy, France \\ $^2$Department of Information Engineering, Universit\`{a} Politecnica delle Marche, Italy}

\begin{document}
%
\maketitle
\begin{abstract}
  Single-channel speech separation has recently made great progress thanks to learned filterbanks as used in ConvTasNet. In parallel, parameterized filterbanks have been proposed for speaker recognition where only center frequencies and bandwidths are learned. In this work, we extend real-valued learned and parameterized filterbanks into complex-valued analytic filterbanks and define a set of corresponding representations and masking strategies. We evaluate these filterbanks on a newly released noisy speech separation dataset (WHAM). The results show that the proposed analytic learned filterbank consistently outperforms the real-valued filterbank of ConvTasNet. Also, we validate the use of parameterized filterbanks and show that complex-valued representations and masks are beneficial in all conditions. Finally, we show that the STFT achieves its best performance for 2~ms windows.\looseness=-1
\end{abstract}
\begin{keywords}
Speech separation, filterbank design.
\end{keywords}
\section{Introduction}
\label{sec:intro}
 Be it for speech intelligibility or automatic speech recognition, speech processing applications need effective speech separation in clean and noisy recording conditions. Single-channel speaker-independent speech separation has recently seen great progress in clean recording conditions. A wide variety of deep learning methods have been introduced \cite{DPCLHershey2016, PITYu2016, DPCL+Isik2016, uPITKolbaek2017,  DANetChen2017, ChimeraWang2018, LSTMLuo2018, FurcaShi2019, ConvLuo2018, CombinedYang2019} and compared on the wsj0-2mix benchmark introduced in \cite{DPCLHershey2016}. All these methods rely on a neural network to estimate the time-frequency mask associated with each source.

 Crucially, the time-frequency transform must allow both signal analysis and resynthesis. It can be either fixed, such as the short time Fourier transform (STFT) \cite{DPCLHershey2016, PITYu2016, DPCL+Isik2016, uPITKolbaek2017,  DANetChen2017, ChimeraWang2018} and its inverse and the Mel \cite{MelWeninger2015} or gammatone \cite{GammaNecciari2018, MultiPhaseDitter2019} filterbanks, or learned jointly with the masking network \cite{LSTMLuo2018, FurcaShi2019, ConvLuo2018, CombinedYang2019}. While learned representations have been shown to be undeniably superior to the STFT for speech separation in clean recording conditions \cite{LSTMLuo2018, FurcaShi2019, ConvLuo2018}, their impact in the presence of noise has been lesser studied. In fact, the authors of \cite{WHAMWichern2019} introduce a noisy extension of wsj0-2mix, WHAM, on which initial results indicate that the advantage of learned representations reduces as noise is introduced, suggesting that learning from the raw waveform might be harder in noisy conditions. In parallel, parameterized kernel-based filterbanks have been introduced as a front-end for speech and speaker recognition \cite{SincNetRavanelli2018, InterpLoweimi2019}. The underlying idea is to restrict the filters to a certain family of functions and jointly learn their parameters with the network. These filterbanks are meant for signal analysis only, though.

 In this paper, we define suitable parameterized filters for analysis-synthesis. Compared with fixed STFT filters, the proposed filters offer more flexibility and diversity thanks to their adaptive center frequency and bandwidth. Conversely, their parameterized form offers fewer parameters to learn and better interpretability compared with their learned counterparts. To do so, we extend the parameterized filters introduced in \cite{SincNetRavanelli2018} to complex-valued analytic filters, thus enabling perfect synthesis via overlap-add and building shift invariance, a desirable property for time-frequency representations. We then propose a similar analytic extension for learned filters. Finally, we evaluate the performance of these analysis-synthesis filterbanks in a unified framework, as a function of window size in both clean and noisy scenarios.

We present the general framework for speech separation and the proposed filterbanks in Section \ref{sec:model}. We describe the experimental setup in Section \ref{sec:setup} and the results in Section \ref{sec:results}. We conclude in Section \ref{sec:concl}.

\section{Model}
\label{sec:model}
Single-channel speech separation is the task of retrieving individual speech sources from a mixture, optionally in the presence of noise. The observed signal $x(t)$ is described as
\begin{equation}
\label{eq:mixture}
x(t) = \sum_{i=1}^C s_i(t) + n(t),
\end{equation}
where $C$ is the number of sources, $\{s_i(t)\}_{i=1..C}$ are the individual source signals and $n(t)$ is additive noise. The task is then to produce accurate estimates $\Test{s_i(t)}$ of each $s_i(t)$.

\subsection{General framework}
\label{ssec:framework}
Most state-of-the-art speech separation methods can be described using an encoding-masking-decoding framework. An encoder transforms the time-domain signal by convolving every signal frame indexed by $k \in \{0,...,K\minus 1\}$ with a bank of $N$ \textit{analysis} filters $\{u_n(t)\}_{n=0..N-1}$ of length $L$:\footnote{Mathematically, this is a correlation rather than a convolution.}
\begin{equation}
\label{eq:analysis}
\Xmat(k, n) = \sum_{t=0}^{L-1} x(t + kH) u_n(t), \quad n \in \{0, ..., N \minus 1 \} ,
\end{equation}
where $H$ is the \textit{hop size}.
After an optional non-linearity $\mathcal{G}$, $\Xmat$ is then fed to the masking network $\mathcal{MN}$ :
\begin{equation}
\label{eq:Mask estimation}
\mathcal{MN}(\mathcal{G}(\Xmat)) = [\Mmat_1, ..., \Mmat_C].
\end{equation}
Each estimated mask $\Mmat_i$ is multiplied with the input to obtain the estimated representation of source $i$:
\begin{equation}
\label{eq:Mask apply}
\Ymat_i = \mathcal{G}(\Xmat) \odot \Mmat_i, \quad i \in \{1,...,C\},
\end{equation}
with $\odot$ denoting point-wise multiplication. 
The decoder maps each $\Ymat_i$ to the time domain by transposed convolution with a bank of $N$ \textit{synthesis} filters $\{v_n(t)\}_{n=0..N-1}$ of length $L$:
\begin{equation}
\label{eq:synthesis}
\Test{s_i(t)} = \sum_{k=0}^{K-1} \sum_{n=0}^{N-1} \Ymat_i(k, n) v_n(t-kH).
\end{equation}
The analysis and synthesis filters fall into three categories: \emph{free}, \emph{parameterized}, or \emph{fixed}. In \cite{LSTMLuo2018, ConvLuo2018, FurcaShi2019}, the filters are free: all weights $\{u_n(t)\}$ and $\{v_n(t)\}$ are jointly learned with the masking network. Parameterized filters belong to a family of filters, whose parameters are jointly learned with the network instead\cite{SincNetRavanelli2018, InterpLoweimi2019}. For instance, the filters in \cite{SincNetRavanelli2018} are defined as the difference between two low-pass filters with cutoff frequencies $f_1$ and $f_2$:
\begin{align}
\nonumber
u_n(t; \theta)& = 2 f_2 \sinc(2\pi f_2 n) - 2 f_1 \sinc(2 \pi f_1 n) \\
\label{eq:sincos prod}
& =  2 f_w \sinc(2\pi f_w n) \cos(2\pi f_c n),
\end{align}
where $\theta=\{f_1, f_2\}$, $f_w=f_2-f_1$, and $f_c=(f_1+f_2)/2$. All filters drawn from this family are even functions, thus making it unsuitable for resynthesis. Finally, fixed filters represent handcrafted transforms such as the STFT \cite{DPCLHershey2016}, gammatone\cite{GammaNecciari2018} or Mel \cite{MelWeninger2015} filters. In the case of the STFT:
\begin{align}
\label{eq:stft filters}
u_n(t) = h_\text{a}(t) e^{-2j\pi n/ N} \enskip \text{and} \enskip v_n(t) = h_\text{s}(t) e^{2j\pi n/ N},
\end{align}
with $h_\text{a}$ and $h_\text{s}$ the analysis and synthesis windows.

A desirable property of time-frequency representations is \textit{shift invariance}, i.e., invariance to small delays in the time domain. \textit{Analytic filters}\cite{AnalyticFlanagan1980} have this property. Namely, the modulus of the convolution between a real-valued signal and an analytic filter is the envelope of that signal in the frequency band defined by the filter. The STFT filters \eqref{eq:stft filters} are examples of such analytic filters, and the magnitude of the STFT is the corresponding shift-invariant representation. Given any real-valued filter $u(t) \in \mathbb{R}^{1 \times L}$, a corresponding analytic filter $u_\text{analytic}(t)$ can be obtained as
\begin{equation}
\label{eq:analytic filt}
u_\text{analytic}(t) = u(t) + j \mathcal{H}[u(t)]
\end{equation}
where $\mathcal{H}$ denotes the Hilbert transform which imparts a $\minus \pi / 2$ phase shift to each positive frequency component. In the following, we detail the proposed analytic expansion of both parameterized and free filters.

\subsection{Proposed analytic filterbanks}
\label{ssec:extending param}
We define parameterized analytic analysis filters $u_n$ as
\begin{align}
\label{eq:analysis sinc}
\nonumber
u_n(t; \theta) &=  2 f_w \sinc(2\pi f_w t) (\cos(2\pi f_c t) - j\sin(2\pi f_c t))\\
&= 2 f_w \sinc(2\pi f_w t) e^{-2j\pi f_c t}.
\end{align}
This complements the original family of even filters \eqref{eq:sincos prod} with odd ones. The new family $\{u_n\}_{n=1..N}$ can form a complete basis of the signal space, and each filter is \textit{analytic} so that
\begin{equation}
\Im(u_n(t; \theta)) = \mathcal{H}[ \Re(u_n(t, \theta))].
\end{equation}
The corresponding family of synthesis filters is defined as:
\begin{equation}
\label{eq:synthesis sinc}
v_n(t; \phi) = 2 g_n f_w \sinc(2\pi f_w t) e^{2j\pi f_c t},
\end{equation}
where $\phi=\{f_1, f_2, g\}$, and g$_n$ is a gain parameter learned to improve resynthesis. Finally, each filter is multiplied by a Hamming window of size $L$, as in \cite{SincNetRavanelli2018, InterpLoweimi2019}.

Similarly, in the case of free filters, we propose to ensure that the learned filters are analytic by parameterizing them by their real part and computing the corresponding analytic filter via \eqref{eq:analytic filt} during the forward pass of the network. This is applied to both analysis and synthesis filters.

In the following, analytic parameterized and free filterbanks are respectively denoted as \textit{param}$+\mathcal{H}$ and \textit{free}$+\mathcal{H}$.

\subsection{Network inputs and output masks}
\label{ssec:inp_mask}
Analytic filterbanks can be viewed either as a set of $N$ complex filters or as $2N$ real filters. This opens different possibilities for the inputs and outputs of the masking network. We consider three possibilities for the input representation: the modulus of $\Xmat$ (\textit{Mag}), its real and imaginary parts (\textit{Re+Im}) or a concatenation of both (\textit{Mag+Re+Im}). The masks can be applied to the modulus of $\Xmat$ (\textit{Mag}) or to its real and imaginary parts using either a complex-valued product with $\Mmat_k \in \mathbb{C}^N$ (\textit{Compl}) or a real-valued product with $\Mmat_k \in \mathbb{R}^{2N}$ (\textit{Re+Im}).

\subsection{Masking network}
\label{ssec:arch}
The masking network is chosen to be the time-domain Convolutional Network (TCN) in \cite{ConvLuo2018}. It comprises $R$ convolutional blocks, each consisting of $X$ 1-D dilated convolutional layers with exponentially increasing dilation factor. In \cite{ConvLuo2018}, the best system used $R=3$ and $X=8$. For most of our experiments, we use a lighter network (\textit{Light TCN}) with $R=2$ and $X=6$ to reduce training time. The systems achieving the best performances with the light TCN are then retrained using the larger network (\textit{Full TCN}). The hop size is set to $H=L/2$. No non-linearity $\mathcal{G}$ is applied to the inputs and the masks are estimated using ReLU as the activation function\footnote{Our implementation can be found at \href{https://github.com/mpariente/AsSteroid}{github.com/mpariente/AsSteroid}.}.
\section{Experimental procedure}
\label{sec:setup}
\subsection{Dataset}
\label{ssec:dataset}
The systems are evaluated on clean (wsj0-2mix \cite{DPCLHershey2016}) and noisy (WHAM \cite{WHAMWichern2019}) two-speaker mixtures created with the scripts in \cite{WHAMweb}. In the clean condition, a 30~h training set and a 10~h validation set are generated by mixing randomly selected utterances from different speakers in the Wall Street Journal (WSJ) training set $si\_tr\_s$ at random signal-to-noise (SNR) ratios between 0 and 5~dB. Noisy datasets are then created by mixing noiseless mixtures with noise samples at SNRs between -3 and 6~dB with respect to the loudest speaker. For both conditions, a 5~h evaluation set is designed similarly with different speakers and noise samples. All Light TCN experiments are conducted with a sampling rate of 8~kHz.

\subsection{Training and evaluation setup}
\label{ssec:training}
Training is performed on 4~s segments using the permutation-invariant \cite{PITYu2016, uPITKolbaek2017} scale-invarariant source-to-distor\-sion ratio (SI-SDR) \cite{ConvLuo2018,SISDRLeroux2019} as the training objective. For Light TCN, Adam \cite{Adam} with an initial learning rate of $1.10^{-3}$ is used as the optimizer. Learning rate halving and early stopping are applied based on validation performance. The best models are retrained with the Full TCN architecture using rectified Adam \cite{RAdamLiu2019} with look ahead \cite{LookAheadZhang2019}. Mean SI-SDR improvement (SI-SDR\textsubscript{i}) is reported for all models on their respective test sets.

\section{Results}
\label{sec:results} 
\subsection{Light TCN experiments}
\label{ssec:Lightexp}
\textbf{Analycity of parameterized and free filterbanks}: We first evaluate the role of analycity in our parameterized filterbank \eqref{eq:analysis sinc}. We consider $N=512$ filters, with \textit{Mag+Re+Im} input and \textit{Mag} mask. To compensate for the greater number of filters, we set $N=1536$ for the non-analytic filterbank \eqref{eq:sincos prod}. Table \ref{table:cos vs cos_sin} shows that the original filterbank is unsuitable for analysis-synthesis for any window size and that the proposed analytic extension overcomes this issue.

\begin{table}[h!] 
	\begin{flushleft} 
	\end{flushleft}
	\centering
	\small
	\begin{tabular}{lrrrrr}
	\toprule
	Window size (ms) &  2  &  5  & 10  &  25 &  50 \\
	\midrule
	Param.			    &   2.3 &   1.0 &  0.6 & -0.8 & -2.7 \\
	Param.(3x filters) &   2.3 &   1.2 &  0.7 & -0.7 & -2.7 \\
	Param.$+\mathcal{H}$ & \textbf{11.8} &  \textbf{11.6} &  \textbf{9.1} &  \textbf{7.3} &  \textbf{4.0} \\
	\bottomrule
	\end{tabular}
    \caption{SI-SDR\textsubscript{i} (dB) as a function of window size for parametric filterbanks in clean conditions. Bold values represent the best statistically significant results.}
\label{table:cos vs cos_sin}
\end{table}

\begin{figure}[h!]
  \centering
  \includegraphics[width=\linewidth]{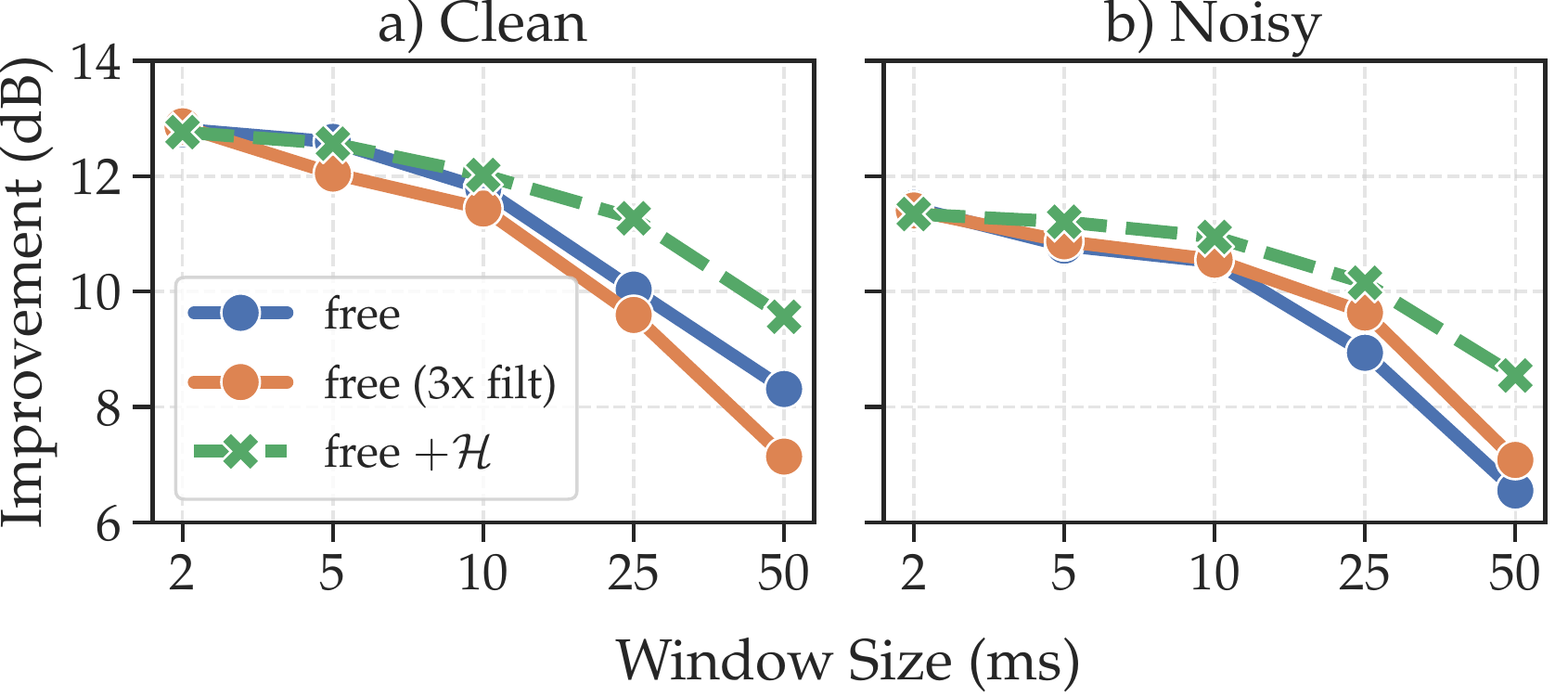}
  \vspace{-0.6cm}
  \caption{SI-SDR\textsubscript{i} as a function of window size for free filterbanks in clean and noisy conditions.}
  \label{fig:tasnet3x}
\end{figure}

The results of a similar experiment for the free filterbanks in both clean and noisy conditions are shown in Fig.~\ref{fig:tasnet3x}. While analytic extension of free filters doesn't hurt performance for short windows, we can gain up to 2~dB for larger windows. Also note that tripling the number of free filters  doesn't match the gain brought by analycity.

\begin{figure*}[]
  \centering
  \includegraphics[width=\linewidth]{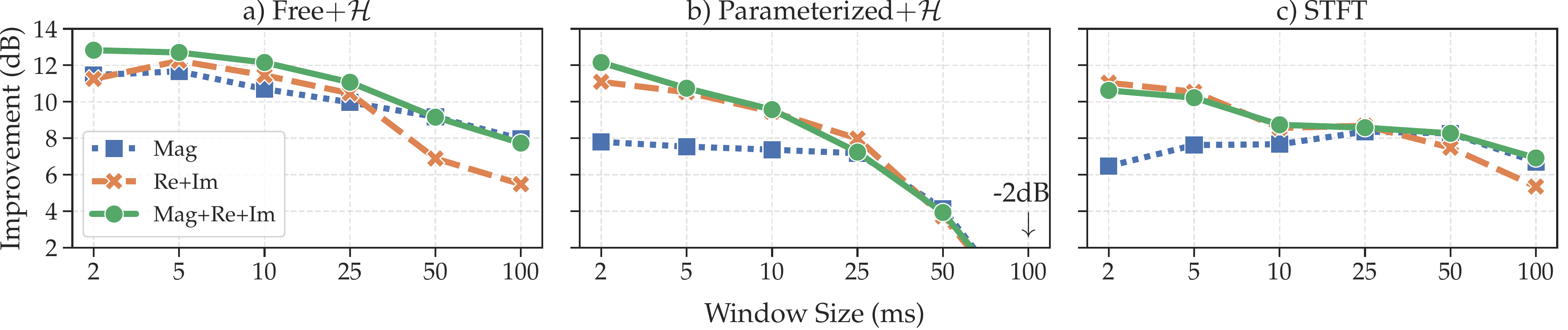}
  \vspace{-0.6cm}
  \caption{SI-SDR\textsubscript{i} for different inputs to the network as a function of window size for a) free$+\mathcal{H}$, b) param.$+\mathcal{H}$ and c) STFT.}
\label{fig:inp_exp}
\end{figure*}

\noindent\textbf{Masking strategies}: Next, we evaluate the choice of the masking strategy for \textit{Mag+Re+Im} input. Table \ref{table:MaskTable} shows that this  has very little and moderate impact on the free$+\mathcal{H}$ and param.$+\mathcal{H}$ filterbanks, respectively. For the STFT, the gap between \textit{Mag} and \textit{Re+Im} masks suggest that phase modeling is indeed necessary for good separation with small windows.

\begin{table}[h]
\centering
\setlength\tabcolsep{4pt} 
\small
	\begin{tabular}{l ccc c ccc}
	\toprule
	& \multicolumn{3}{c}{Clean}&& \multicolumn{3}{c}{Noisy} \\
	Filterbank & Mag & Compl & Re+Im && Mag & Compl & Re+Im\\
	
	\midrule
	Free$+\mathcal{H}$   & 12.7 & 12.8 & 12.8 & & 11.0 & 11.3 & 11.0 \\
	Param.$+\mathcal{H}$ & 11.8 & 12.2 & 12.5 & & 10.5 & 10.6 & 10.1 \\
	STFT                 & 9.8  & 10.5 & 10.9 & & 9.4 & 9.4  & 9.9 \\
	\bottomrule
	\end{tabular}
	\caption{SI-SDR\textsubscript{i} (dB) as a function of the mask type for each analytic filterbank in clean and noisy conditions, $L=16$.}
\label{table:MaskTable}
\end{table}

\noindent\textbf{Input representations}: Next, we evaluate the impact of the input representation as a function of window size. The results are plotted in Fig.~\ref{fig:inp_exp} for the \textit{Re+Im} mask. In the case of free$+\mathcal{H}$, the shift-invariant representation \textit{Mag} helps only for large windows. For param.$+\mathcal{H}$, the \textit{Re+Im} representation is sufficient for all window sizes. Finally, for the STFT, the \textit{Mag} and \textit{Re+Im} inputs complement each other so that maximum performance is reached with \textit{Re+Im} input for small windows, when phase modeling is necessary, and with \textit{Mag} for larger windows when amplitude modeling is sufficient.

\begin{figure}[h]
  \centering
  \includegraphics[width=\linewidth]{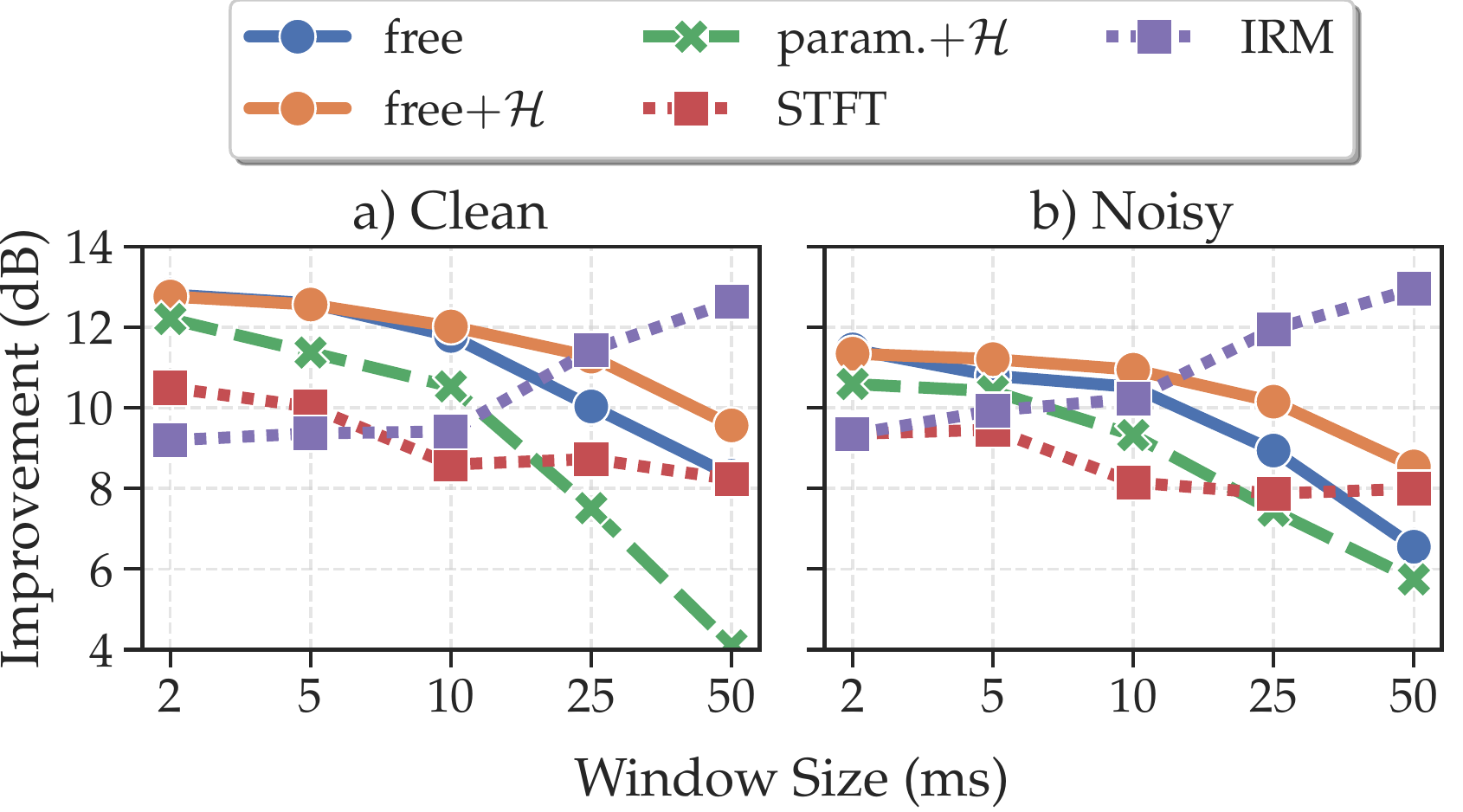}
  \vspace{-0.6cm}
  \caption{SI-SDR\textsubscript{i} as a function of window size for all filterbanks and for the ideal ratio mask (IRM).}
  \label{fig:main_plot}
\end{figure}

\noindent\textbf{Filterbank choice}: Finally, we compare the original free filterbank and all analytic filterbanks in Fig.~\ref{fig:main_plot}, in both clean and noisy conditions. We use \textit{Mag+Re+Im} input and \textit{Re+Im} masking for all methods based on analytic filters.

The take-away messages of the above experiments are as follows. First, parameterized filters as in \cite{SincNetRavanelli2018} are unsuitable for separation. The proposed analytic extension addresses this issue but performance decreases as the window size increases. Second, analytic extension of learned filters stabilizes performance for large windows. Third, combining complex inputs and masks for small windows brings the best results for all analytic filterbanks. Interestingly, this also holds for the STFT.

\subsection{Full TCN experiment}
We retrained the two best models, i.e., free with $L=16$ (a.k.a.~Conv-TasNet) and free$+\mathcal{H}$ with $L=16$, with the full TCN in clean and noisy conditions and for both the 8~kHz \textit{min} and 16~kHz \textit{max} versions of the dataset. The results are reported in Table \ref{table:FinalResults} along with the chimera++\cite{ChimeraWang2018} results in \cite{WHAMWichern2019}. Compared to Conv-TasNet, the proposed analytic extension improves the results in all tested conditions by up to 0.7~dB, showing that shift-invariant representations can benefit Conv-TasNet's TCN even for small windows.

\begin{table}[h] 
	\centering	
	\setlength\tabcolsep{4pt}
	\small
	\begin{tabular}{ll  c c}
	\toprule
	Model &  Dataset  & separate-clean & separate-noisy \\
	\midrule
	\midrule
	chimera++ \cite{ChimeraWang2018}  & 8kHz min       &11.0      &9.9  \\
	Conv-TasNet \cite{ConvLuo2018}\footnote{Our implementation.}  & 8kHz min       &  15.1      &  \textbf{12.7} \\
	Free$+\mathcal{H}$&  8kHz min    &  \textbf{15.8}      &  \textbf{12.9} \\
	\midrule
	chimera++ \cite{ChimeraWang2018} &   16kHz max     &9.6       &10.2  \\
	Conv-TasNet \cite{ConvLuo2018} &   16kHz max     & 13.6   &  13.3 \\
	Free$+\mathcal{H}$&  16kHz max  &  \textbf{14.0}  &  \textbf{14.0} \\
	\bottomrule
	\end{tabular}
        \caption{SI-SDR\textsubscript{i} (dB) comparison between the proposed analytic free filterbank and previously proposed models. Bold values represent the best statistically significant results.}
\label{table:FinalResults}
\end{table}

\section{Conclusion}
\label{sec:concl}
In this paper, we defined analytic extensions of both parameterized and free filterbanks. The resulting filterbanks are more interpretable, and perform equally well or better than their real-valued counterparts for speech separation in clean or noisy conditions. Final evaluation with the most expressive TCN from \cite{ConvLuo2018} showed that using analytic filterbanks consistently improved separation performances over all conditions. Also, to the best of our knowledge, this is the first time parameterized filterbanks are used in an end-to-end speech separation framework. Although they don't perform as well as free filterbanks, we argue that a better design could bridge this gap. Finally, we showed that complex-valued inputs and masks can be beneficial for separation with short windows for all filterbanks. In particular, the implicit phase modeling capability of TasNet's TCN applies to the STFT, which achieves its best performance for 2~ms windows. 

\bibliographystyle{IEEEbib}
\bibliography{refs}

\end{document}